\documentclass[twocolumn,showpacs,preprintnumbers,amsmath,amssymb]{revtex4}

\usepackage{graphicx}
\usepackage{dcolumn}
\usepackage{bm}

\begin{document}

\title{Intrinsic detection efficiency of superconducting single photon detector in the modified hot spot model}

\author{A.N. Zotova$^{1,2}$}
\email{vodolazov@ipmras.ru}
\author{D.Yu. Vodolazov$^{1,2}$,}
\affiliation{$^1$ Institute for Physics of Microstructures,
Russian Academy of Sciences, 603950,
Nizhny Novgorod, GSP-105, Russia \\
$^2$ Lobachevsky State University of Nizhni Novgorod, 23 Gagarin
Avenue, 603950 Nizhni Novgorod, Russia}

\begin{abstract}

We theoretically study the dependence of the intrinsic detection
efficiency (IDE) of superconducting single photon detector on the
applied current $I$ and magnetic field $H$. We find that the
current, at which the resistive state appears in the
superconducting film, depends on the position of the hot spot
(region with suppressed superconductivity around the place where
the photon has been absorbed) with respect to the edges of the
film. It provides inevitable smooth dependence IDE(I) when IDE
$\sim 0.05-1$ even for homogenous straight superconducting film
and in the absence of fluctuations. When IDE $\lesssim 0.05$ much
sharper current dependence comes from the fluctuation assisted
vortex entry to the hot spot located near the edge of the film. We
find that weak magnetic field strongly affects IDE when the photon
detection is connected with fluctuation assisted vortex entry
(IDE$\ll 1$) and it weakly affects IDE when the photon detection
is connected with the current induced vortex entry to the hot spot
or nucleation of the vortex-antivortex pair inside the hot spot
(IDE$\sim 0.05-1$).
\end{abstract}

\maketitle

\section{Introduction}

At the moment there are several phenomenological models which try
to describe the detection mechanism of superconducting single
photon detectors  - SSPD (for their comparison see for example
Ref. \cite{Lusche_mechanism}). In this paper we study how one of
the main characteristic of SSPD - intrinsic detection efficiency
(IDE) depends on the current and magnetic field in the modified
hot spot model, which takes into account both the nonuniform
current distribution around the hot spot (current crowding effect
\cite{Clem_crowding}) and suppression of the superconductivity by
the current \cite{Zotova}. In comparison with the system detection
efficiency, which defines the probability to detect the photon by
the whole detector, IDE has a meaning of probability to detect the
photon when it is absorbed by the main element of SSPD -
superconducting current-carrying film in the form of meander
\cite{Hofherr}.

The used modified hot spot model has two main qualitative
differences with previous hot spot models
\cite{Semenov_first,Maingault,Engel_JAP,Semenov_hot_spot}: we
solve the current continuity equation $div j=0$ in the film with
the hot spot (which automatically gives us the maximal value of
the current density near the hot spot even if it is located far
from the edges of the film) and we take into account the back
effect of the current redistribution on the superconducting order
parameter in the film with hot spot (which gives us nucleation of
the vortex-antivortex pair inside the hot spot located far from
the edges of the film). The used phenomenological model of the hot
spot cannot relate quantitatively the energy of the photon with
the size of the hot spot and how strong the superconductivity is
suppressed inside it. But we demonstrate that the radius of the
hot spot and level of suppression of superconductivity affect the
photon detection mechanism and dependencies IDE(I) and IDE(H) only
quantitatively.

When the hot spot (region with partially or completely suppressed
superconductivity) appears in the superconducting film after
photon absorption the superconducting current avoids that region
and it 'crowds' near the hot spot. The current crowding effect in
thin superconducting films with geometric inhomogeneities
attracted large attention last years both from theoretical
\cite{Clem_crowding,Clem_corner} and experimental
\cite{Henrich,Hortensius,Akhlaghi,Adami} points of view. In Ref.
\cite{Clem_crowding} among different problems there was considered
the superconducting semi-infinite film with the edge defect in the
form of the semicircle (semicircular notch - see Fig. 18 in Ref.
\cite{Clem_crowding}). This problem is mathematically equivalent
to the infinite film with a circular notch and when the diameter
of the notch $D=2R$ is much larger than the coherence length the
critical current is equal to the half of the depairing current
(see Fig. 19(b) in Ref. \cite{Clem_crowding}) which is consequence
of the current crowding near the notch. This result coincides with
our result for the finite width film with the normal spot (see Eq.
(12) in Ref. \cite{Zotova} in the limit $\gamma \to 0$) when $w\gg
2R$, i.e. when the film becomes formally 'infinite'. Finite width
of the film changes this universal result and besides the critical
current depends now on the position of the normal spot (or notch)
with respect to the edges of the film.

In our previous work \cite{Zotova} we consider two locations of
the hot spot: in the center and at the edge of the film. We find
that in these two limiting cases the resistive state starts (in
absence of fluctuations) at different critical currents, which
were called as a detection currents $I_{det}$, via appearance of
the current induced vortices and their motion across the film. We
stress here that vortex nucleation is {\it the direct consequence}
of the spatially nonuniform distribution of the superconducting
order parameter and supercurrent in the superconducting film with
the hot spot.

In this work we find $I_{det}$ at different locations of the hot
spot in the film. We show that $I_{det}$ reaches the minimal value
$I_{det}^{min}$ when the hot spot is located near the edge of the
film and it has maximal value when the hot spot is located in the
center of the film. We also calculate the energy barrier for the
vortex entry to the superconducting film with the hot spot when
$I<I_{det}^{min}$ and find the rate of fluctuation assisted vortex
entry as function of the current. These results allow us to
calculate dependence IDE(I) in wide range of the currents and we
argue that vortices play important role both when IDE$\simeq 1$
and when IDE$\ll 1$. Application of the magnetic field decreases
locally the current density in one half of the film and increases
it in another half because of field induced screening currents. We
show that in the used model it leads to strong increase of IDE
when the photon detection is governed by the fluctuation assisted
vortex entry to the film (IDE$\ll 1$) and the effect becomes much
weaker when the photon detection is determined by the current
induced vortex entry (IDE$\sim 0.1-1$).

\section{Model}

In our hot spot model (as in other hot spot models
\cite{Semenov_first,Maingault,Engel_JAP,Semenov_hot_spot}) it is
assumed that in the place, where the photon is absorbed, there is
nonequilibrium ('heated') distribution of the quasiparticles over
the energy which locally suppresses the superconductivity and it
leads to redistribution of the current density in the film.
Calculation of actual nonequilibrium distribution function of
quasiparticles $f_{neq}$ needs solution of the kinetic equation
and it is very complicated problem \cite{Kozorezov} and we do not
study it in this work. Our aim is to study how the presence of the
region with locally suppressed superconductivity (superconducting
order parameter) affects the value of the critical current, at
which the superconducting state of the film with hot spot (HS)
becomes unstable. For this purpose we numerically solve the
Ginzburg-Landau equation for the superconducting order parameter
\begin{equation}
\xi_{GL}^2\left( \nabla -i\frac{2e}{\hbar
c}A\right)^2\Delta+\left(1-\frac{T_{bath}}{T_c}+\Phi_1-\frac{|\Delta|^2}{\Delta_{GL}^2}\right)\Delta=0
\end{equation}
with the additional term
\begin{equation}
\Phi_1=\int_{|\Delta|}^{\infty}\frac{2(f^0-f)}{\sqrt{\epsilon^2-|\Delta|^2}}d\epsilon,
\end{equation}
which takes into account the impact of nonequilibrium
quasiparticle distribution function $f(\epsilon)\neq
f^0(\epsilon)=1/(exp(\epsilon/k_BT_{bath})+1)$. In Eq. (1) $A$ is
a vector potential, $\xi_{GL}^2=\pi\hbar D/8k_BT_c$ and
$\Delta_{GL}^2=8\pi^2(k_BT_c)^2/7\zeta(3)\simeq 9.36(k_BT_c)^2$
are the zero temperature Ginzburg-Landau coherence length and the
order parameter correspondingly. Together with Eq. (1) we also
solve continuity equation $div j_s=0$ ($j_s$ is a superconducting
current density) to find the distribution of the current density.

In numerical calculations it is convenient to use dimensionless
units. Therefore we scale the length in units
$\xi(T_{bath})=\xi_{GL}/(1-T_{bath}/T_c)^{1/2}$, $\Delta$ in units
$\Delta_{eq}=\Delta_{GL}(1-T_{bath}/T_c)^{1/2}$ and $A$ in units
$\Phi_0/2\pi \xi$ ($\Phi_0$ is a magnetic flux quantum). In these
units Eq. (1) has a following form
\begin{equation}
\left(\nabla-i{\tilde
A}\right)^2{\tilde\Delta}+\left(\alpha-|{\tilde\Delta}|^2\right){\tilde\Delta}=0,
\end{equation}
where $\alpha=(1-T_{bath}/T_c+\Phi_1)/(1-T_{bath}/T_c)$.

In Eq. (3) the effect of absorbed photon on the superconducting
properties of the film is described by the parameter $\alpha$ (in
equilibrium $\alpha=1$) which is determined by $f(\epsilon)$. In
our model we put $\alpha=const<1$ inside the hot spot region which
leads to local suppression of $|\Delta|$ not only inside but also
outside the hot spot (due to proximity effect). Surely this
assumption oversimplifies the real situation where $\alpha$
depends on the coordinate and we cannot expect that our results
are {\it quantitatively} valid. But below we demonstrate that {\it
qualitatively} the obtained results does not depend on actual
value of $\alpha$ (which governs the suppression of $|\Delta|$
inside HS) and we expect that they are valid in the real situation
with coordinate dependent $\alpha(r)$.

To have an insight on the possible values of $\alpha$ one can use
the local temperature approach which implies that $f(\epsilon)$
can be described by the Fermi-Dirac function with the local
temperature $T_{loc}$ which is different from the bath temperature
$T_{bath}$. It is easy to show (with help of Eq. (3)) that in this
limit
\begin{equation}
\alpha(\vec{r},t)=(1-T_{loc}(\vec{r},t)/T_c)/(1-T_{bath}/T_c)
\end{equation}
The area where $T_{loc}>T_{bath}$ increases in time due to
diffusion of hot quasiparticles from the place where the photon
was absorbed. From Eq. (3) it follows that the order parameter is
suppressed stronger in the place where $T_{loc}>T_c$ and
$\alpha<0$. At some moment in time the region where $T_{loc}>T_c$
reaches the maximal size and it is naturally to model the hot spot
by the circle with radius $R$ and put $\alpha=0$ inside the
circle. In this case the radius of the spot $R$ and energy of the
absorbed photon $ch/\lambda$ are roughly related as
\begin{equation}
\eta \frac{ch}{\lambda}\simeq d\pi R^2\frac{H_{cm}^2}{8\pi}
\end{equation}
where $H_{cm}=\Phi_0/2\sqrt{2}\pi\xi\lambda_L^2$ is the
thermodynamic magnetic field, $\lambda_L$ is the London
penetration depth, $d$ is the thickness of the film and
$H_{cm}^2/8\pi$ is the superconducting condensation energy per
unit of volume. Coefficient $0<\eta<1$ takes into account that
only part of the energy of the photon is delivered for suppression
of $\Delta$ and the rest of the photon's energy goes for heating
of quasiparticles and phonons.

When the photon is absorbed at the edge of the film the
nonequilibrium quasiparticles cannot leave the sample and we model
the hot spot by the semicircle with larger radius $R'=\sqrt{2}R$
to keep the area of the hot spot unchanged.
\begin{figure}[hbtp]
\includegraphics[width=0.48\textwidth]{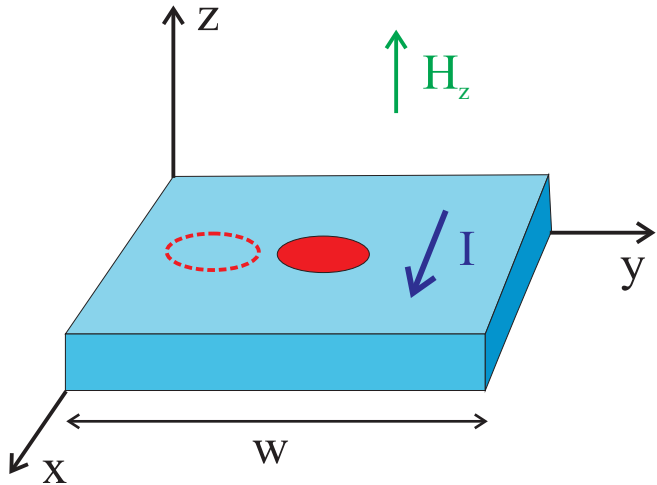}
\caption{Model geometry: two-dimensional film with width $w$ and
transport current $I$ placed in perpendicular magnetic field. The
hot spot is modelled by the finite size region where $\alpha<1$
(it models the heating of quasiparticles in this region due to
absorbed photon).}
\end{figure}

From Eq. (3) it follows that for the spots with $R\gg \xi$ the
order parameter inside the hot spot is $\simeq
\sqrt{\alpha}\Delta_{eq}$ when $\alpha\geq 0$, and it is equal to
zero when $\alpha<0$. Remind here that different $\alpha$
corresponds to different level of the nonequilibrium inside the
hot spot (in the local temperature approximation this relation is
given by Eq. (4)). Due to proximity effect the order parameter is
suppressed partially also at $r>R$ and it becomes larger than
$\sqrt{\alpha}\Delta_{eq}$ inside the hot spot when the radius is
about of the coherence length.

In numerical calculations we consider the film of finite width $w$
and length $L=4w$ with different locations of the hot spot (region
where $\alpha<1$) across the film - see Fig.1. We also add to the
right hand side of Eq. (3) the term with time derivative $\partial
{\tilde \Delta}/\partial t$ which allows us to find not only the
value of the critical current (above this current there is no
stationary solution of Eq. (3)) but also the place in the film
where vortices nucleate.

\section{Detection current}

Let us now to discuss what is the mechanism of destruction of the
superconducitng state in the superconducting film with the photon
induced hot spot. We find that when the hot spot is located at the
edge of the film and the current exceeds the critical value
$I_{pass}$ the vortex enters the hot spot via edge of the film and
than it passes through the film (see the sketch in the Fig. 2(a)).
In Ref. \cite{Zotova}, in the local temperature approach, we find
that the vortex motion may strongly heat the superconductor and it
leads to the appearance of the normal domain. In the following we
assume that passage of even single vortex through the film is a
sufficient condition for destruction of the superconducting state
in the superconductor biased at the current, not much smaller than
the depairing current.

We also find that for relatively large radius of the hot spot
($R\gtrsim 3\xi$ when $\alpha=0$) at currents $I_{en}<I<I_{pass}$
the vortex enters the hot spot, but it cannot leave it (the
similar effect was found earlier in Ref. \cite{Vodolazov_defect}
at study of the effect of edge defects on the vortex penetration
to the superconducting film). The hot spot, as a region with
suppressed superconductivity, could be considered as a photon
induced pinning center and vortex becomes unpinned only at the
current $I \geq I_{pass}$. But in contrast with the usual pinning
center the hot spot exists only during short period of time $min
(\tau_{e-ph},\tau_{e-e})$, where $\tau_{e-ph}$ and $\tau_{e-e}$
are the inelastic relaxation times due to electron-phonon or
electron-electron interaction, respectively. Indeed, the
nonequilibrium quasiparticles with energy $\epsilon<\Delta_{eq}$
cannot diffuse out the hot spot and they relax to equilibrium via
(whatever is stronger) electron-phonon or electron-electron
interaction with quasiparticles having energy
$\epsilon>\Delta_{eq}$ and which can diffuse away from the region
with suppressed $|\Delta|$.
\begin{figure}[hbtp]
\includegraphics[width=0.5\textwidth]{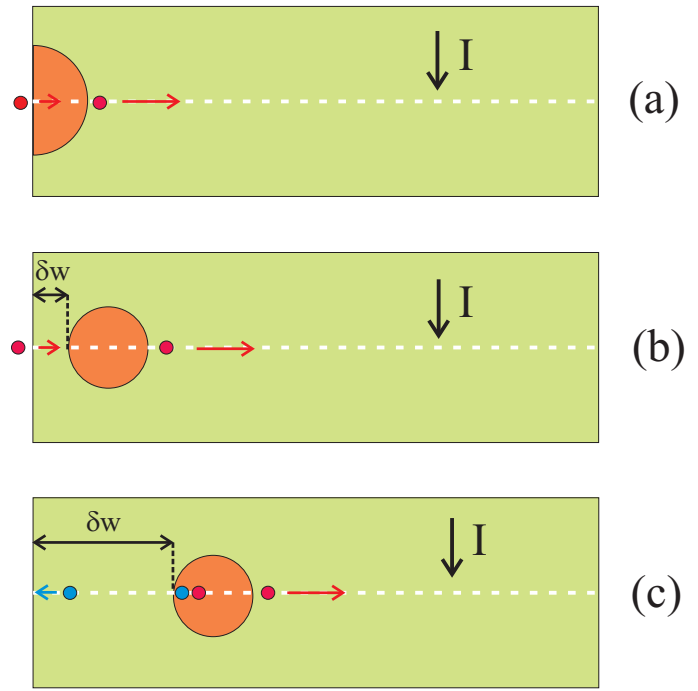}
\caption{Schematic representation of mechanisms of destruction of
the superconducting state at different locations of the hot spot:
(a) the hot spot in the form of semicircle is located at the edge
of the film, (b) the hot spot is located close to the edge of the
film ($\delta w \lesssim 2 \xi$), c) the hot spot is located at
distance $\delta w \gtrsim 2\xi$. In cases (a,b) the vortex enters
via the nearest edge of the film, while in the case (c) the
vortex/antivortex pair is nucleated inside the hot spot.}
\end{figure}

Therefore, in the range of the currents $I_{en}<I<I_{pass}$ the
vortex is temporarily pinned and after 'dissociation' of the hot
spot it becomes unpinned and can pass through the film or exit via
the nearest edge due to interaction with its 'image' outside the
film. Our numerical simulations with help of time dependent
Ginzburg-Landau equation and time dependent $\alpha(t)$ confirm
both scenario. Starting from the hot spot state with $\alpha=0$
and pinned vortex we gradually increase $\alpha$ in time up to its
equilibrium value $\alpha=1$ (no hot spot state). We find that
when the current is just above $I_{en}$ the vortex exit via the
nearest edge, while at relatively larger currents (less than
$I_{pass}$) vortex goes through the film.

When the spot is located near the edge of the film (see Fig. 2(b))
there are the same characteristic currents $I_{en}$ and $I_{pass}$
- first one corresponds to the vortex entrance to the hot spot and
the second one to its unpinning and the free passage of the vortex
through the film. In contrast with the case drawn in Fig. 2(a),
when the hot spot 'dissociates' the vortex passes the film in the
whole current interval $I_{en}<I<I_{pass}$. We explain it by the
larger distance from the nearest edge and smaller attraction force
from the 'image' of the vortex.

When the distance ($\delta w$ in Fig. 2(b,c)) between the 'edge'
of hot spot and the edge of the film exceeds $\sim 2 \xi$ the
vortex/antivortex pair is nucleated {\it inside} the hot spot at
the current $I_{pair}$. Again, if $R\gtrsim 3 \xi$ vortex and
antivortex becomes unpinned at larger current $I_{pass}>I_{pair}$.
For such a location the 'dissociation' of the hot spot leads to
the annihilation of the vortex-antivortex pair and their passage
through the film occurs only at $I \geq I_{pass}$.

We find that with increasing the radius of the hot spot the gap
between currents $I_{en}$ (or $I_{pair}$) and $I_{pass}$
increases. For spots with $R \lesssim 3 \xi$ and $\alpha=0$ these
two currents coincide (which is connected with relatively large
value of $|\Delta|$ inside the spot and worse ability to pin the
vortices). The obtained results are rather general and depend on
the specific value of $\alpha$ and width of the film only
quantitatively.
\begin{figure}[hbtp]
\includegraphics[width=0.53\textwidth]{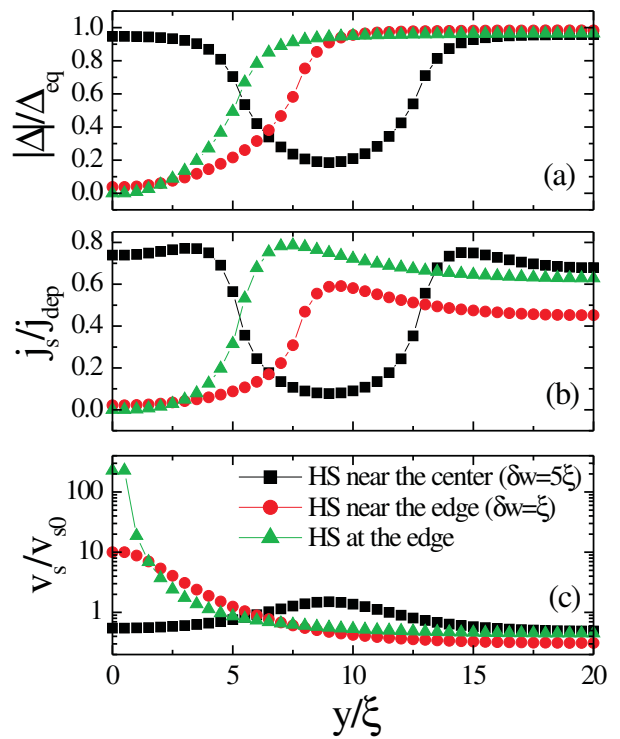}
\caption{Distribution of the order parameter (a), current density
(b) and supervelocity (c) across the film (along the dashed lines
shown in the Figs. 2(a-c)) with different locations of the hot
spot. The applied current is just below $I_{en}$(or $I_{pair}$).}
\end{figure}

The physical origin for the found results is following. Due to the
current crowding effect the current density reaches maximal value
near the hot spot (see Fig. 3(b)). But simultaneously there is an
enlargement of the supervelocity $v_s\sim j_s/|\Delta|^2$ in or
around the hot spot (see Fig. 3(c)). The last effect is crucial
for stability of the superconductivity because the superconducting
state becomes unstable when the velocity of superconducting
electrons exceeds some critical value \cite{Vodolazov_defect}
(which is similar to the instability criteria for flowing
superfluid helium). In the system with the uniform distribution of
the order parameter and current density it coincides with the
condition that the current density reaches the depairing current
density $j_{dep}$. But inside and around the hot spot there is a
gradient of both $|\Delta|$ and $j$ (see Figs. 3(a,b)), these two
criteria do not coincide and the vortices nucleate in the place
where the supervelocity reaches the maximal value. Note, that
quantitatively both these criteria are not too far from each
other, because when the vortex enters the hot spot (or the
vortex/antivortex pair is nucleated inside the hot spot) the
maximal current density near the spot is close to $j_{dep}$ (see
Fig. 3(b)).

With above findings we define the photon detection current
$I_{det}$ as a current at which at least one vortex can pass
through the film after appearance of the photon induced hot spot.
This current is equal to $I_{pass}$ when the vortex/antivortex
pair is nucleated inside the hot spot or it is equal to $I_{en}$
when the single vortex enters the hot spot via edge of the film.
For the hot spot located at the edge of the film (see Fig. 2(a))
we also take into account that the vortex passes the film at
current a bit larger than $I_{en}$.

\section{Dependence of $I_{det}$ on the location of the hot spot at H=0}

In Fig. 4 we present dependence of $I_{det}$ on the coordinate of
center of HS with different radiuses $R=2\xi$, $4\xi$ and $5 \xi$
which roughly correspond to the photons with $\lambda/\eta= 25 \mu
m$, $6.3 \mu m$ and $4.0 \mu m$, respectively (note, that
$\eta\simeq 0.1-0.4$ according to previous estimations
\cite{Semenov_hot_spot,Engel_TaN} and for calculation of $\lambda$
with help of Eq. (5) we use parameters of TaN film
\cite{Engel_TaN} with thickness $d=3.9 nm$). One can see that the
minimum of the detection current is reached when HS 'touches' the
edge of the film and it is maximal when HS seats in the center of
the film. We argue that this result is the consequence of
different current crowding at different locations of the hot spot
in the film. Indeed, when HS approaches the edge of the film the
current crowding increases in the narrowest sidewalk (and
simultaneously the supervelocity inside the hot spot increases)
and the vortices nucleate and becomes unpinned at the smaller
applied current. But when the hot spot approaches the edge of the
film its linear size decreases from $2R$ up to $\sqrt{2}R$ (we
keep the area of the hot spot constant, because it is created by
the photons of the same energy) and the current crowding
decreases. It leads to increasing the applied current at which the
vortex can enter the hot spot.
\begin{figure}[hbtp]
\includegraphics[width=0.53\textwidth]{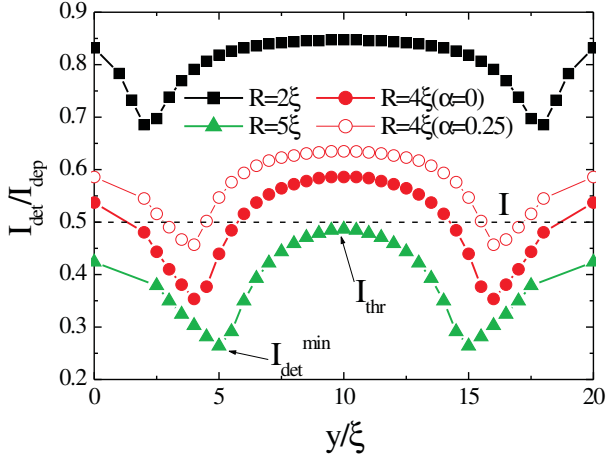}
\caption{Dependence of the detection current on the location of
the hot spot with different radiuses in the film with $w=20 \xi$}.
\end{figure}

How the results shown in Fig. 4 are related to the intrinsic
detection efficiency of superconducting single photon detectors?
Consider for example the photon which creates the hot spot with
radius $R=4\xi$. Let the transport current $I$ is equal to $0.5
I_{dep}$ (dashed line in Fig. 4). Then the part of the film where
$I_{det}<I$ detects the absorbed photons while the rest of the
film (regions near both edges and the center of the film) cannot
detect such a photons and IDE$<1$. Only when the current exceeds
the threshold value ($I_{thr}=I_{det}^{max}\simeq 0.58 I_{dep}$
for $R=4\xi$) the whole film participates in detection of photons
and IDE$=1$.
\begin{figure}[hbtp]
\includegraphics[width=0.44\textwidth]{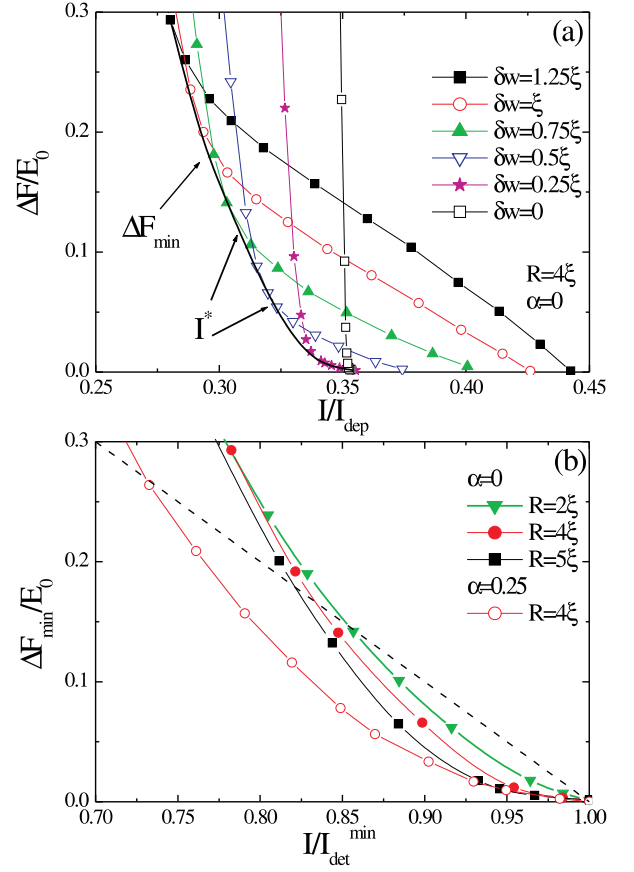}
\caption{(a) Energy barrier for the vortex entry to the hot spot
with radius $R=4\xi$ ($\alpha=0$) located at different distances
$\delta w$ from the edge of the film. Solid line shows the minimal
energy barrier at given value of the current. (b) Current
dependence of the minimal barrier for the vortex entry to the hot
spots with different radiuses and $\alpha$. Dashed line
corresponds to the dependence $\Delta F/E_0=1-I/I_c$ following
from the London model for the vortex entry to the straight film
without the hot spot (when $I\sim I_c$).}
\end{figure}

If the transport current $I<I_{det}^{min}\simeq 0.35 I_{dep}$ then
IDE goes to zero in absence of fluctuations (when $R=4\xi$).
Fluctuations favor the creation of the vortices and they may
provide the finite IDE even at $I<I_{det}^{min}$. To distinguish
this process from the current induced vortex penetration at the
current $I>I_{en}$ we use the term 'fluctuation induced vortex
penetration' when $I<I_{en}$. Because the barrier for the vortex
entry increases rapidly with decreasing current
\cite{Bartolf,Bulaevskii,Vodolazov_dU} the main contribution to
the fluctuation assisted IDE$\neq 0$ comes from the photons which
create the hot spot near the edge of the film, where $I_{det}$ is
minimal. For this location of HS the vortex enters via the edge of
the film and one needs to calculate the energy barrier for the
vortex entry to the hot spot at $I<I_{en}=I_{det}$. In the Fig.
5(a) we show calculated barrier $\Delta F$ (the energy is scaled
in units of $E_0=\Phi_0^2d/16\pi^2\lambda_L^2$) for the different
locations of the hot spot with $R=4\xi$ ($\Delta F$ is found using
the numerical procedure from Ref. \cite{Vodolazov_dU}). The energy
barrier increases rapidly below some current $I^*(\delta
w)<I_{det}(\delta w)$ (see Fig. 5(a)) because at currents
$I\lesssim I^*$ the vortex cannot be pinned by the hot spot and it
exits via the nearest edge of the film.

One can see from the Fig. 5(a) that for given value of the current
there is a minimal barrier for the vortex entry $\Delta F_{min}$
when the hot spot is located at specific distance from the edge of
the film. In Fig. 5(b) we show dependence $\Delta F_{min}(I)$
found for different radiuses of HS and $\alpha$. Note that when
$I\sim I_{det}^{min}$ the energy barrier increases much slowly
with the current decrease as compared to the dependence $\Delta
F(I)$ for the vortex entry to the film without HS, which follows
from the London model \cite{Bartolf,Bulaevskii} ($\Delta
F_L/E_0\sim (1-I/I_{dep})$ - see dashed line in Fig. 5(b)) or
Giznburg-Landau model \cite{Vodolazov_dU} ($\Delta F_{GL}/E_0\sim
A(w)(1-I/I_{dep})$, with $A(w)\sim 1.5-1.8$ for films with $w=7-30
\xi$) models at $I\sim I_{dep}$. Physically it is connected with
suppressed order parameter in the sidewalk between the hot spot
and edge of the film (see Fig. 3(a)). As a result it costs less
energy for creation of the vortex (or vortex nucleus
\cite{Vodolazov_dU}) there.
\begin{figure}[hbtp]
\includegraphics[width=0.44\textwidth]{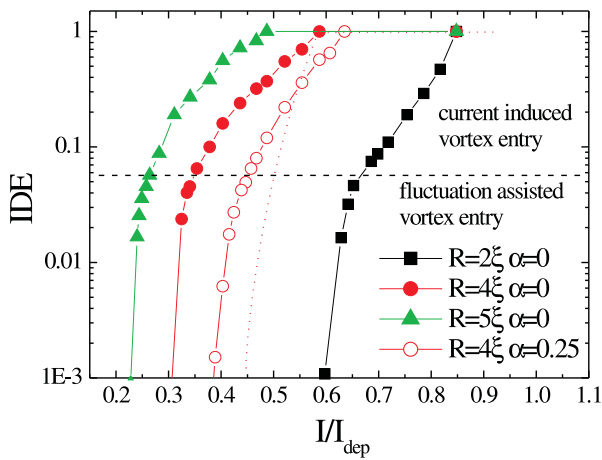}
\caption{Dependence of intrinsic detection efficiency on applied
current which follows from nonmonotonic dependence $I_{det}(y)$
(at $I>I_{det}^{min}$) and finite probability for vortex entry due
to fluctuations at $I<I_{det}^{min}$ (area below the dashed line).
Dotted curve corresponds to dependence IDE(I) following from the
hot belt model \cite{Bulaevskii} for one energy of the photon
(qualitative presentation).}
\end{figure}

IDE in the fluctuation region could be found with help of
Arrhenius law IDE$=\beta exp(-\Delta F_{min}/k_BT)$ where the
coefficient $\beta$ in front of the exponent is equal IDE at
$I\simeq I_{det}^{min}$ (we choose $\beta=0.05$ because of rapid
decrease of $I_{det}$ near the $I_{det}^{min}$ - see Fig. 4).
Using parameters of TaN film \cite{Engel_TaN} ($\lambda_L=560 nm$,
$d=3.9 nm$) and $T=4 K$ we find $F_0/k_BT\simeq 62$. In Fig. 6 we
plot IDE as a function of the current for photons with different
wavelengths (which create the hot spots with different radiuses).
In the same figure we plot the sketch of dependence IDE(I) which
follows from the hot belt model \cite{Bulaevskii} (dotted curve).
In the hot belt model IDE$<1$ is explained exclusively by the
effect of fluctuations, which leads to the fast drop of IDE with
current decrease. Much smoother change of IDE from 1 up to $\simeq
0.05$ in the modified hot spot model exists even at $T=0$ and it
changes on much faster decay when IDE$\lesssim 0.05$ where it is
finite only due to fluctuations (qualitatively it is similar to
the hot belt model).
\begin{figure}[hbtp]
\includegraphics[width=0.5\textwidth]{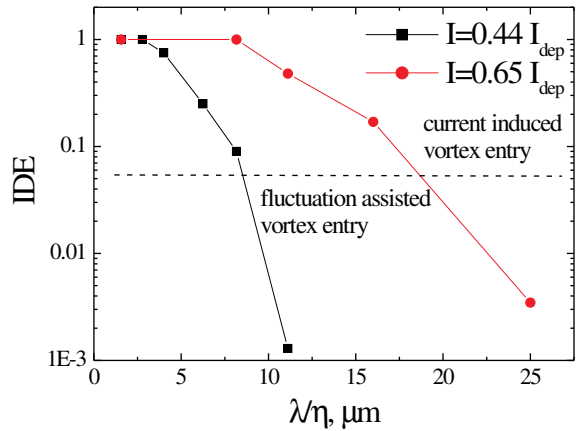}
\caption{Dependence of intrinsic detection efficiency on the
wavelength for the superconducting film with $w=20 \xi$ and
different transport currents. In calculations we use parameters of
TaN film from Ref. \cite{Engel_TaN}.}
\end{figure}

We also calculate dependence IDE$(\lambda)$ at fixed current. For
this purpose one needs to find the part of the film where
$I_{det}(y)$ is smaller than the transport current for chosen $R$
(i.e. wavelength). Because we know the barrier for the vortex
entry to the hot spot we are able to calculate fluctuation induced
IDE too. In Fig. 7 we present results of these calculations. As in
dependence IDE(I) one may distinguish two regions: relatively
smooth variation of IDE with $\lambda$ when it varies in the range
$\sim 0.05-1$ and much faster decay of IDE at larger wavelengths,
where IDE is finite only due to fluctuation assisted vortex entry
to the hot spot. Qualitatively found results resemble
experimentally observed dependence IDE$(\lambda)$ (see for example
Refs. \cite{Hofherr,Engel_TaN}).

\section{Effect of the magnetic field}

How do dependencies $I_{det}(y)$ and IDE(I) change in the presence
of the applied magnetic field? In Fig. 8 we plot the current
density distribution in the superconducting film with and without
perpendicular magnetic field when there is no hot spot. One can
see that in the presence of low magnetic field the current density
increases in the left half of the film and it decreases in the
right half of the film (for opposite direction of H the situation
is opposite). Here under low magnetic field we mean fields
$H<H_s/2$, where $H_s\simeq \Phi_0/4\pi \xi w$ is a magnetic field
at which the surface barrier for vortex entry to the straight
superconducting film is suppressed \cite{Stejic} (for film with
$w=20 \xi$, $H_s/2\simeq 0.025 H_{c2}$). Using the results of
section IV one may expect that the detection current becomes
smaller (in comparison with the case $H=0$) for HS appearing in
the part of the film with locally enhanced current density and
vise versus in the opposite case. Our numerical calculations
support this idea (see Fig. 9). Indeed, when $H$ increases
$I_{det}$ decreases in the left half of the film and it increases
in the right half of the film. Note that $I_{det}$ changes much
weaker in the central part of the film because of relatively small
change of the current density there (see Fig. 8).
\begin{figure}[hbtp]
\includegraphics[width=0.4\textwidth]{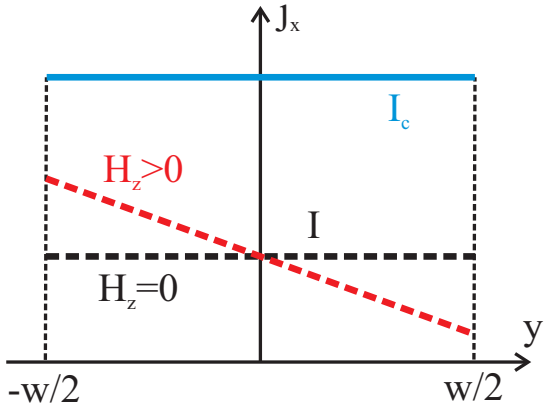}
\caption{Distribution of the current density in the narrow
superconducting film with the transport current, placed in
perpendicular magnetic field.}
\end{figure}

We also calculate the minimal energy barrier for the vortex entry
(see Fig. 10) to the hot spot located near the left and right
edges of the film at different magnetic fields. From Fig. 10 it
follows that the shape of dependence $\Delta F_{min}(I)$ weakly
changes at low magnetic fields $H\ll H_s$ while detection current
in the left and right minima ($I_{det}^{L,R}$) varies linearly
with the magnetic field (see inset in Fig. 10).
\begin{figure}[hbtp]
\includegraphics[width=0.52\textwidth]{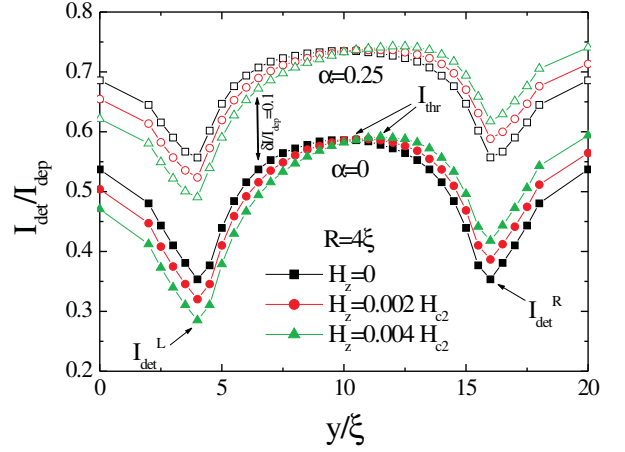}
\caption{Dependence of the detection current on the coordinate of
the hot spot with $R=4\xi$ (and $\alpha=0$, $0.25$) at different
magnetic fields. Data for the hot spot with $\alpha=0.25$ are
shifted upwards by $\delta I/I_{dep}=0.1$ for clarity.}
\end{figure}

With the help of found results we calculate how IDE changes at
weak magnetic fields (see Fig. 11). Because $I_{thr}$ stays
practically unchanged (see Fig. 9) while $I_{det}^{min}$ decreases
with increase of $H$ (see inset in Fig. 10), the strongest change
of IDE occurs at $I<I_{det}^{min}$ when IDE$ \lesssim 0.05$. Note
that the hot belt model \cite{Bulaevskii} predicts relatively
large change of IDE in whole range of $0<$IDE$<1$ and linear
decrease of the threshold current when $H$ increases (see
dash-dotted curves in Fig. 11). Indeed, in the hot belt model
$I_{thr}$ is equal to the critical current of the film with the
hot belt, which decreases linearly at weak magnetic field (like as
$I_{det}^L$ in the inset in Fig. 10).
\begin{figure}[hbtp]
\includegraphics[width=0.52\textwidth]{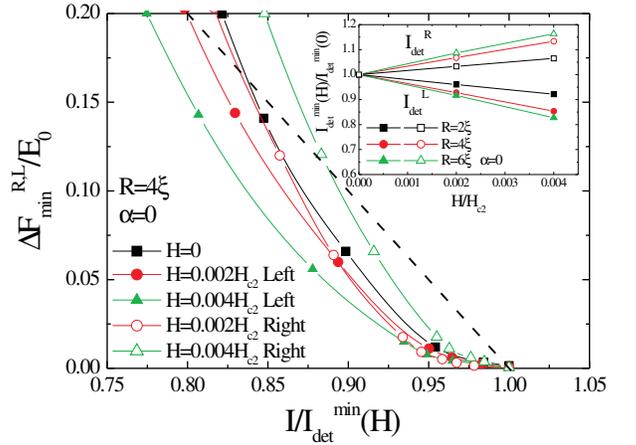}
\caption{Dependence of the minimal energy barrier for the
vortex/antivortex entry to the hot spot located near the
left/right edge of the film at different magnetic fields. Dashed
line: $\Delta F_{min}/E_0=1-I/I_{det}^{min}(H)$. In the inset we
show dependence of $I_{det}^{L,R}(H)$ at low magnetic fields.}
\end{figure}

\section{Discussion}

The vortex-assisted mechanism of the photon detection was
discussed previously in several works
\cite{Zotova,Hofherr,Bulaevskii,Semenov_PhysC}. In contrast with
Refs. \cite{Hofherr,Semenov_PhysC} we argue that the vortices play
important role in all range of $0<$IDE$<1$ and not only when
IDE$\ll 1$ is determined by the fluctuation assisted vortex entry
or unbinding of the vortex-antivortex pair. From the hot belt
model \cite{Bulaevskii} it follows that threshold current
$I_{thr}$ (at which IDE is about of unity) {\it decreases}
linearly with increase of the magnetic field while we predict very
weak dependence (increase) of $I_{thr}$ at low magnetic fields $H
\lesssim H_s \simeq \Phi_0/4\pi \xi w$.

Our model also predicts that at current $I\gtrsim I_{det}^{min}$
(where $I_{det}^{min}$ depends on the radius of the hot spot and,
hence, on the energy of the photon - see Fig. 4) the photon count
rate varies with the magnetic field much weaker than at smaller
currents (see Fig. 11 - note, that some signs of this effect were
observed in Ref. \cite{Lusche_PRB} - see Fig. 3 there). To observe
this effect experimentally it is preferable to use the materials
with the threshold current $I_{thr}$ much smaller than critical
current of the superconducting film (like in materials studied in
Refs. \cite{Engel_TaN,Marsili,Korneeva,Verma}) because in this
case one can vary magnetic field in wide range and do not overcome
the critical current $I_c(H)$. Experimentally $I_{det}^{min}$ for
each photon's wavelength could be determined from the dependence
of detection efficiency (DE) on the current if it saturates (and
DE(I) has a plateau) at large currents. According to our
calculations at $I=I_{det}^{min}$ the DE$\simeq
0.05$DE$_{plateau}$ (which corresponds to IDE $\simeq 0.05$). But
in our model we consider only straight homogenous film, while real
SSPD are based on the superconducting meanders which have the
bends, structural defects and variations of the thickness and/or
width. Therefore the minimal detection current may correspond to
the different location of the hot spot when the found for the
straight film. To demonstrate this effect in Fig. 12 we show
calculated dependence $I_{det}(y)$ for the film with $90^0$ degree
bend (see inset in Fig. 12). From Fig. 12 one can see that the
photon absorbed near the inner corner of the bend (the current
density is maximal there due to current crowding) could be
detected at the smallest transport current. From comparison of
Fig. 12 with Fig. 9 one can also see that the bend acts like a
weak magnetic field. This similarity is not accident because in
both cases there are places in the film where the local current
density is maximal and the minimal detection current corresponds
to the absorption of the photon near that places.
\begin{figure}[hbtp]
\includegraphics[width=0.44\textwidth]{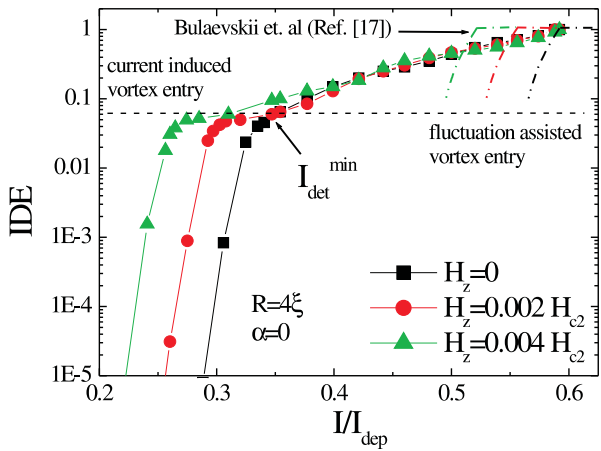}
\caption{Dependence of the intrinsic detection efficiency on the
current at different magnetic fields. Dashed-dotted lines
correspond to the dependences IDE(I) following from the hot belt
model at different magnetic fields \cite{Bulaevskii} (qualitative
presentation).}
\end{figure}

This result demonstrates that in the bent {\it homogenous} film
the area near the bend determines both $I_{det}^{min}$ and minimal
IDE, which is not connected with the fluctuation assisted vortex
entry. For example, at $I=0.28 I_{dep}$ (which is a little above
the minimal detection current for the hot spot with $R=4\xi$
located near the bend - see Fig. 12) the minimal energy barrier
for the vortex entry to the straight part of the film with the hot
spot is about $0.29 E_0$ (see Fig. 5(a)). Taking into account that
typically $E_0/k_BT \gtrsim 50$ one easily find that the vortex
penetration to the straight part of the film is suppressed by
factor $exp(-\Delta F/k_BT )\lesssim 10^{-6}$ which is much
smaller than the ratio between the area near the bends and the
rest of the superconducting meander (which is about
$10^{-2}-10^{-3}$ depending on how to estimate the 'active' area
near the bend). Therefore we expect that in real SSPD the
fluctuation assisted vortex entry contributes to the finite IDE
when it becomes smaller than $\lesssim 10^{-3}-10^{-2}$.

The actual boundary between the fluctuation assisted and the
current induced vortex penetration (leading to the detection of
the photon) could be deduced from the experiment with the magnetic
field. Indeed, because the first mechanism is more sensitive to
the magnetic field (see Fig. 11) the IDE at currents
$I<I_{det}^{min}$ should increase much faster than at larger
currents.
\begin{figure}[hbtp]
\includegraphics[width=0.52\textwidth]{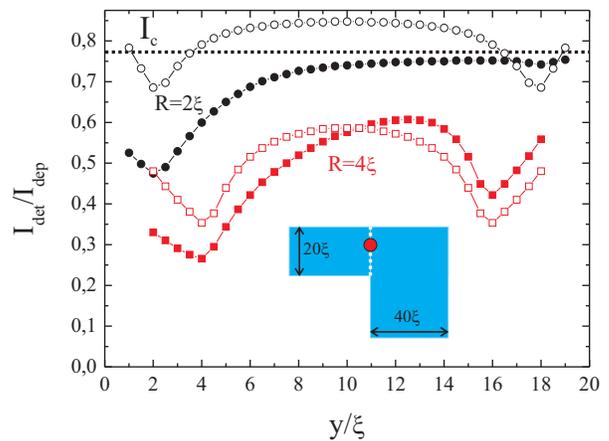}
\caption{Dependence of the detection current on the coordinate of
the hot spot with radius $R=2 \xi$ (solid circles) and $R=4\xi$
(solid squares) along the white line in the bent film shown in the
inset. In the same figure we show $I_{det}$ for the straight film
with $w=20 \xi$ (empty circles and squares). The critical current
of the film with the bend $I_c=0.77 I_{dep}$ (dotted line).}
\end{figure}

Discussed above effect could explain the absence of dependence of
the photon count rate (PCR) on the magnetic field, experimentally
found in Ref. \cite{Engel_PRB}. Indeed, in that work the field
dependence was studied in the current interval where PCR was
changed from its maximal value by two orders of magnitude (which
is equivalent to similar change in IDE). Therefore it might be
that the minimal used current still was larger than
$I_{det}^{min}$. As a result at used in Ref. \cite{Engel_PRB}
magnetic fields $H<100 Oe$ and $H_s\sim 4000 Oe$ (it is called as
$H^*$ in Ref. \cite{Engel_PRB}) the PCR could change not more than
by several percents (following the change of $I_{det}^{min}$ - see
Fig. 9), which is about of the experimental error in Ref.
\cite{Engel_PRB}.

On the contrary, in Ref. \cite{Lusche_PRB} the strong dependence
of PCR on magnetic field was found at low currents. Comparison of
the obtained results with the predictions of the hot belt model
\cite{Bulaevskii} made in Ref. \cite{Lusche_PRB} demonstrated good
qualitative but bad quantitative agreement. To fit the
experimental results authors needed special dependence of the
characteristic vortex energy (in the presence of the hot belt -
see Eq. (3) in Ref. \cite{Lusche_PRB}) on the wavelength and
current (see insets in Fig. 2(b) and 3 in Ref. \cite{Lusche_PRB})
which does not follow from the theory of Ref. \cite{Bulaevskii}.
Our model also predicts the strong dependence of PCR on the
magnetic field, but only at the current
$I<I_{det}^{min}(\lambda)$. Because the current dependence of the
energy barrier for the vortex entry is not linear (see Figs.
5(b),10) we expect quasi-exponential increase of PCR at low
magnetic fields which differs from the exponential law (see for
example Eq. (2) in \cite{Lusche_PRB} or Eq. (5) in
\cite{Engel_PRB}) following from the linear dependence $\Delta
F(I)$ in the London model. To make the quantitative comparison
with Ref. \cite{Lusche_PRB} one needs to calculate the IDE(I) for
the film with the bends (using the same procedure as in the
present work for the straight film) and find the energy barrier
for the vortex entry to the hot spot, located close to the bend.
Our present results give only qualitative prediction that there
exist some current $I_{det}^{min}$ (at this current IDE$\sim
10^{-3}-10^{-2}$) above which the photon count rate weakly depends
on H and at smaller currents it depends on weak magnetic fields
much stronger (quasi-exponentially).

\section{Conclusion}

In the framework of the modified hot spot model we predict that
the intrinsic detection efficiency of superconducting single
photon detector gradually changes with the current. The change of
IDE from 1 to $\sim 0.01$ occurs due to dependence of the current,
at which the resistive response appears, on the location of the
hot spot in the film. The resistive state starts from the vortex
entry (when the hot spot is located near or at the edge of the
film) or nucleation of the vortex-antivortex pair (when hot spot
is located far from the edge) and their motion across the film.
The change of IDE from $\sim 0.01$ up to zero is connected with
the fluctuation assisted vortex entry to the hot spot located near
the edge of the film, when the current itself cannot cause the
vortex entry.

Weak applied magnetic field ($H\ll \Phi_0/4\pi\xi w$ strongly
affects (increases) IDE$\ll 1$ when it is nonzero only due to
fluctuation assisted vortex entry to the hot spot. At the currents
close to the threshold current, at which IDE $\simeq 1$, the
applied magnetic field weakly affects the detection ability and
may provide only increase of $I_{thr}$.

\begin{acknowledgments}

We thank to Alexander Semenov, Gregory Gol'tsman, Alexey Semenov,
Jelmer Renema, Michiel de Dood and Martin van Exter for valuable
discussion of the found results. The work was partially supported
by the Russian Foundation for Basic Research (project 12-02-00509)
and by the Ministry of education and science of the Russian
Federation (the agreement of August 27, 2013,  N 02.Â.49.21.0003
between The Ministry of education and science of the Russian
Federation and Lobachevsky State University of Nizhni Novgorod).

\end{acknowledgments}



\begin{references}

\bibitem{Lusche_mechanism} R. Lusche, A. Semenov, H. Huebers, K. Ilin, M. Siegel,
Y. Korneeva, A. Trifonov, A. Korneev, G. Goltsman, and D.
Vodolazov, arXiv:1303.4546.

\bibitem{Zotova} A. Zotova and D. Y. Vodolazov, Phys. Rev. B {\bf 85}, 024509 (2012).

\bibitem{Hofherr} M. Hofherr, D. Rall, K. Il'in, M. Siegel, A. Semenov, H.-W. H\"{u}bers, and
N. A. Gippius, J. of Appl. Phys. {\bf 108}, 014507 (2010).

\bibitem{Semenov_first} A. D. Semenov, G. N. Gol'tsman, and A. A. Korneev, Phys. C
(Amsterdam) {\bf 351}, 349 (2001).

\bibitem{Maingault} L. Maingault, M. Tarkhov, I. Florya, A. Semenov, R. Espiau de Lama\"{e}stre,
P. Cavalier, G. Gol'tsman, J.-P. Poizat, and J.-C. Vill\'{e}gier,
J. Appl. Phys. {\bf 107}, 116103 (2010).

\bibitem{Semenov_hot_spot} A. Semenov, A. Engel, H.-W. H\"{u}bers, K. Il'in, and M. Siegel,
Eur. Phys. J. B {\bf 47}, 495 (2005).

\bibitem{Engel_JAP} A. Engel and A. Schilling, J. Appl. Phys. {\bf 114}, 214501
(2013).

\bibitem{Clem_crowding} J. R. Clem and K. K. Berggren, Phys. Rev. B {\bf 84}, 174510 (2011).

\bibitem{Clem_corner} J. R. Clem, Y. Mawatari, G.R. Berdiyorov and F.M. Peeters, Phys. Rev. B {\bf 85}, 144511
(2012).

\bibitem{Henrich} D. Henrich, P. Reichensperger, M. Hofherr, K. Il'in,
M. Siegel, A. Semenov, A. Zotova and D. Yu. Vodolazov, Phys. Rev.
B {\bf 86}, 144504 (2012).

\bibitem{Hortensius} H. L. Hortensius, E. F. C. Driessen, T. M. Klapwijk,
K. K. Berggren and J. R. Clem, Appl. Phys. Lett. {\bf 100}, 182602
(2012).

\bibitem{Akhlaghi} M. K. Akhlaghi, H. Atikian, A. Eftekharian , M. Loncar,
A. H. Majedi, Optics Express, {\bf 20}, 23610 (2012).

\bibitem{Adami} O.-A. Adami, D. Cerbu, D. Cabosari, M. Motta, J. Cuppens, W.A.
Ortiz, V.V. Moshchalkov, B. Hackens, R. Delamare, J. Van de
Vondel, A.V. Silhanek, Appl. Phys. Lett. 102, 052603, 2013

\bibitem{Kozorezov} A. G. Kozorezov, A. F. Volkov, J. K. Wigmore,
A. Peacock, A. Poelaert, and R. den Hartog, Phys. Rev. B {\bf 61},
11 807 (2000).

\bibitem{Vodolazov_defect} D.Y. Vodolazov, I.L. Maksimov, E.H.
Brandt, Physica C {\bf 384}, 211 (2003).

\bibitem{Engel_TaN} A. Engel, A. Aeschbacher, K. Inderbitzin, A. Schilling, K.
Il'in, M. Hofherr, M. Siegel, A. Semenov, and H.-W. H\"{u}bers,
Appl. Phys. Lett. {\bf 100}, 062601 (2012).

\bibitem{Bartolf} H. Bartolf, A. Engel, A. Schilling, K. Il'in, M. Siegel, H.-W.
H\"{u}bers and A. Semenov, Phys. Rev. B {\bf 81}, 024502 (2010).

\bibitem{Bulaevskii} L. N. Bulaevskii, M. J. Graf, and
V. G. Kogan, Phys. Rev. B {\bf 85}, 014505 (2012).

\bibitem{Vodolazov_dU} D. Y. Vodolazov, Phys. Rev. B {\bf 85}, 174507 (2012).

\bibitem{Stejic} G. Stejic, A. Gurevich, E. Kadyrov, D. Christen, R. Joynt, and
D.C. Larbalestier, Phys. Rev. B {\bf 49}, 1274 (1994).

\bibitem{Semenov_PhysC} A. D. Semenov, P. Haas , H.-W. H\"{u}bers, Konstantin Il'in,
M. Siegel, A. Kirste, T. Schurig, A. Engel, Physica C {\bf 468}
627 (2008).

\bibitem{Marsili} F. Marsili, V. B. Verma, J. A. Stern, S. Harrington, A. E.
Lita, T. Gerrits, I. Vayshenker, B. Baek, M. D. Shaw, R. P. Mirin,
and S. W. Nam, Nature Photonics, {\bf 7}, 210, 2013.

\bibitem{Korneeva} Y. P. Korneeva, M. Y. Mikhailov, Y. P. Pershin, N. N. Manova,
A. V. Divochiy, Y. B. Vakhtomin, A. A. Korneev, K. V. Smirnov, A.
Y. Devizenko, and G. N. Goltsman, arXiv:1309.7074
[cond-mat.supr-con].

\bibitem{Verma} V. B. Verma, A. E. Lita, M. R. Vissers, F. Marsili, D. P. Pappas, R. P. Mirin, and S. W.
Nam, arXiv:1402.4526 [cond-mat.supr-con].

\bibitem{Lusche_PRB} R. Lusche, A. Semenov, Y. Korneeva, A. Trifonov, A. Korneev, G. Gol'tsman,
and H.-W. H\"{u}bers, Phys. Rev. B {\bf 89}, 104513 (2014).

\bibitem{Engel_PRB} A. Engel, A. Schilling, K. Il'in and M. Siegel, Phys. Rev. B {\bf 86},
140506(R) (2012).

\end{references}
\end{document}